\documentstyle[epsfig,multicol,pre,aps]{revtex}
\begin{document}
\draft

\title{Dynamic Stabilization in The Double-Well Duffing Oscillator}

\author{Sang-Yoon Kim${}^{1,}$
\footnote{Electronic address: sykim@cc.kangwon.ac.kr}
 and Youngtae Kim${}^{2,}$
\footnote{Electronic address: ytkim@madang.ajou.ac.kr}
       }
 \address{
 ${}^1$Department of Physics, Kangwon National University,
 Chunchon, Kangwon-Do 200-701, Korea \\
 ${}^2$ Department of Physics, Ajou University,
 Suwon, Kyunggi-Do 442-749, Korea
 }

\maketitle

\begin{abstract}
Bifurcations associated with stability of the saddle fixed point of 
the Poincar\'{e} map, arising from the unstable equilibrium point of 
the potential, are investigated in a forced Duffing oscillator with 
a double-well potential. One interesting behavior is the dynamic 
stabilization of the saddle fixed point. When the driving amplitude 
is increased through a threshold value, the saddle fixed point 
becomes stabilized via a pitchfork bifurcation. We note that this 
dynamic stabilization is similar to that of the inverted pendulum 
with a vertically oscillating suspension point. After the dynamic 
stabilization, the double-well Duffing oscillator behaves as the 
single-well Duffing oscillator, because the effect of the central 
potential barrier on the dynamics of the system becomes negligible.  
\end{abstract}

\pacs{PACS numbers: 05.45.-a}


\begin{multicols}{2}
A periodically driven double-well Duffing oscillator 
\cite{DWD}, which has become a classic model for analysis of 
nonlinear phenomena, is investigated. It can be described 
in a normalized form by a second-order nonautonomous ordinary 
differential equations, 
\begin{equation}
\ddot{x} + \gamma \dot{x} - x + x^3 = A \cos{\omega t},
\label{eq:DDO1}
\end{equation}
where $\gamma$ is the  damping coefficient, and $A$ and $\omega$
are the amplitude and frequency of the external driving force, 
respectively.

The Duffing equation (\ref{eq:DDO1}) with negative linear stiffness 
describes the dynamics of a buckled beam \cite{Holmes,Moon1} 
as well as a  plasma oscillator \cite{Mahaffey}. Its regular and 
chaotic dynamics has been analyzed in great details by Holmes using 
both the theoretical techniques and the computer simulations 
\cite{Holmes}. The results of this work have been also confirmed in 
experiments by Moon \cite{Moon1} for a buckeled elastic beam. 
Since then, a number of authors have studied the forced double-well 
Duffing oscillator in the past two decades, and found rich dynamical 
behaviors \cite{DWD} such as the fractal basin boundary between 
coexisting competing attractors \cite{Moon2}, hopping cross-well 
chaotic state \cite{Hop}, and so on.

Here we are interested in the dynamical behaviors associated with the
saddle fixed point of the Poincar\'{e} map, arising from the unstable
equilibrium point of the potential. One interesting behavior, 
associated with chaotic dynamics, is the homoclinic intersection of 
the stable and unstable manifolds of the saddle fixed point. In 
Ref.~\cite{Holmes}, Holmes showed that as the forcing amplitude $A$ 
is increased the stable and unstable manifolds intersect transversally, 
giving rise to homoclinic motions. However, as $A$ increases further, 
another interesting behavior, associated with stability of the saddle 
fixed point, occurs. When $A$ passes through a threshold value, the 
saddle fixed point becomes stabilized via a pitchfork bifurcation. 
We note that the best-known example of this dynamic stabilization is 
the inverted pendulum with a vertically oscillating suspension point 
\cite{IP}. To our knowledge, this is the first report on such a 
dynamic stabilization in the double-well Duffing oscillator. After the 
dynamic stabilization, the behaviors of the double-well Duffing 
oscillator closely resembles those of the single-well Duffing 
oscillator \cite{Parlitz1}, because the central barrier of the 
potential has no significant effect on the motion of the system. 

For the numerical calculations we transform the second-order ordinary 
differential equation (\ref{eq:DDO1}) into a system of two first-order 
ordinary differential equations:
\begin{equation}
{\dot x} = y,~
{\dot y} = - \gamma y + x -x^3 + A \cos \omega t.
\label{eq:DDO2}
\end{equation}
These equations have a symmetry $S$, because the transformation
\begin{equation}
S: x \rightarrow -x,~y \rightarrow -y,~ t  \rightarrow 
t + {T \over 2}\;\;[T(\rm{period})= {2 \pi \over \omega}],
\label{eq:S}
\end{equation}
leaves Eq.~(\ref{eq:DDO2}) invariant. If an orbit $z(t) [\equiv 
(x(y),y(t))]$ is invariant under $S$, it is called a symmetric orbit. 
Otherwise, it is called an asymmetric orbit and has its ``conjugate'' 
orbit $S z(t)$.

For the unforced case of $A=0$, there exist a saddle equilibrium
point at $(x,y)=(0,0)$ and a conjugate pair of stable equilibrium
points at $(x,y)=(\pm 1,0)$. However, as $A$ is increased from $0$, 
one symmetric saddle-type orbit and two asymmetric attracting orbits 
with the same period ${2 \pi} / \omega$ arise from the saddle 
equilibrium point and the two stable equilibrium points, respectively
\cite{Holmes}. We note that they become the fixed points of the 
Poincar\'{e} map $P$, generated by stroboscopically sampling the 
orbit points with the external driving period $T$. Hereafter we will 
denote the symmetirc saddle fixed point and the asymmetric stable 
fixed points by $z^*_{\rm s}$ and $z^*_{\rm a}$, respectively. Here 
we are interested in the bifurcations associated with stability of 
the symmetric saddle fixed point. Its linear stability is determined 
from the eigenvalues, called the Floquet multipliers, of the 
linearized Poincar\'{e} map $DP$, which can be obtained using the 
Floquet theory \cite{Gucken1}. 

Since the Poincar\'{e} map $P$ with a symmetry $S$ has a constant 
Jacobian determinant (det) less than unity (det $DP = e^{-\gamma T})$, 
the only possible bifurcations of periodic orbits are saddle-node 
(SN), pitchfork (PF), and period-doubling (PD) bifurcations
\cite{Gucken2}. When a Floquet multiplier passes through $1$, a PF or 
SN bifurcation takes place. On the other hand, when it passes through 
$-1$, a PD bifurcation occurs. Hopf bifurcations are excluded. 

Each bifurcation curve in the parameter plane will be classified by 
a pair $(p,q)$ invariant along the curve. Here $q$ denotes the 
period of the orbit and $p$ denotes the torsion number, which 
characterizes the average rotation number of the nearby orbits during 
the period $q$ \cite{Parlitz2}. The torsion number (normalized by the 
factor $2 \pi$) at the PF or SN bifurcation curve becomes an integer. 
However, when crossing a PD bifurcation curve, not only the period 
but also the torsion number doubles from an odd multiple of $1/2$ to 
an odd integer. To keep $p$ as an integer, we choose the pair of 
$(p,q)$ for the PD bifurcation as that of the period-doubled orbit. 
We also note that the SN bifurcations may create symmetric and 
asymmetric orbits with the same $(p,q)$. To differentiate them, 
their symmetries are also used to classify the SN bifurcations. For 
the case of the symmetric (asymmetric) orbit, a letter ``s'' (``a'') 
will be added in the third entry such as $[p,q,$ ``s'' (``a'')] to 
label the corresponding ``symmetric'' (``asymmetric'') SN bifrucation 
curve.

\begin{figure}
\centerline{\epsfig{file={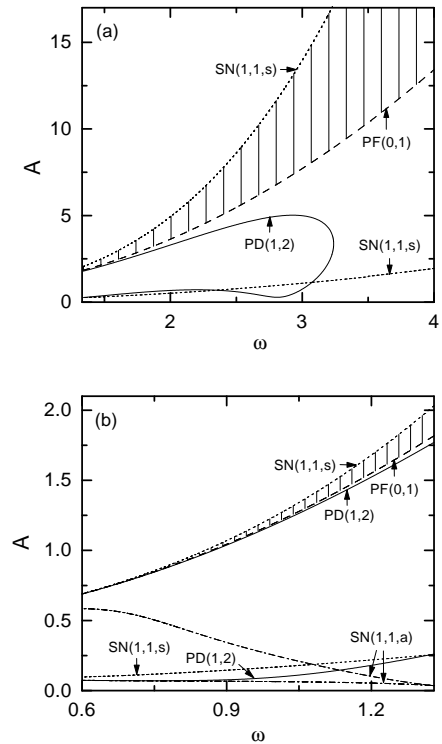}, width=5.7cm}}
\vspace{0.5cm}
\caption{Stability diagram of the symmetric saddle fixed point
 for (a) $\omega > \omega_{\rm r} (\simeq \sqrt{2})$ and (b) $\omega 
 < \omega_{\rm r}$. The hatched region with vertical lines is just its  
 stability region. The symbols SN, PF, and PD denote the saddle-node, 
 pitchfork, and period-doubling bifurcation curves, respectively. Each 
 curve is also labelled by a pair $(p,q)$; $p$ and $q$ are the torsion 
 number and period, respectively. To differentiate the symmetric and 
 asymmetric SN bifurcations, the letters ``s'' and ``a'' are also 
 added in the third entry such as [p,q,``s''(``a'')]. For other 
 details, see the text.
     }
\label{fig:SD}
\end{figure}

By varying the two parameters $A$ and $\omega$, we numerically 
investigate the bifurcation behavior associated with stability of the 
symmetric saddle fixed point for a moderately damped case of 
$\gamma = 0.1$. The associated bifurcation structure in the $\omega-A$ 
plane is shown in Fig.~\ref{fig:SD}. Note that the region, hatched with 
vertical lines, is just the stability region of the symmetric saddle 
fixed point. It is bounded by a lower PF bifurcation curve PF(0,1), 
denoted by a dashed curve, and by an upper symmetric SN bifurcation 
curve SN(1,1,s), denoted by a dotted curve. When crossing the lower PF 
bifurcation curve PF(0,1), the saddle fixed point becomes 
stabilized through a PF bifurcation by absorbing a 
pair of asymmetirc fixed points. As a result of this dynamic 
stabilization, a symmetric stable orbit with period $ {2 \pi} / 
\omega$ (fixed point for the Poincar\'{e} map), encircling the 
unstable equilibrium point of the potential, appears. Note that this 
symmetric stabilized orbit corresponds to the symmetric stable orbit, 
arising from the stable equilibrium point of the potential in the 
single-well Duffing oscillator. Hence the dynamical behavior after 
such a dynamical stabilization becomes essentially the same as that 
of the single-well Duffing oscillator \cite{Parlitz1}. Such stabilized 
symmetric fixed point disappears at the upper symmetric SN bifurcation 
curve SN(1,1,s) by ab-

\noindent{sorbing a symmetric unstable fixed point, born at 
the lower symmetric SN bifurcation curve SN(1,1,s).}

\vspace{-0.4cm}
For the unforced and undamped case, locally the Duffing oscillator 
near the two stable equilibrium points at $(x,y)=(\pm 1,0)$ behaves as 
a soft spring with the natural frequency $\sqrt{2}$. Hence the main 
resonance occurs at $\omega =\omega_{\rm r} (\simeq \sqrt{2})$ in 
the linear limit. For $\omega > \omega_{\rm r}$ the symmetric saddle 
fixed point $z^*_{\rm s}$ becomes stabilized at the PF bifurcation 
curve PF(0,1) in Fig.~\ref{fig:SD}(a) by absorbing a pair of 
asymmetric stable fixed points $z^*_{\rm a}$. However, for $\omega 
< \omega_{\rm r}$, the asymmetric stable fixed points $z^*_{\rm a}$ 
disappear at the upper asymmetric SN bifurcation curve SN(1,1,a), 
denoted by a dash-dotted curve in Fig.~\ref{fig:SD}(b), through the 
collision with the asymmetric unstable fixed points born at the lower 
asymmetric SN bifurcation curve SN(1,1,a). After that, the symmetric 
saddle fixed point $z^*_{\rm s}$ becomes stable at the PF bifurcation 
curve PF(0,1) by absorbing  a pair of asymmetric stable fixed 
points born at the lower asymmetric SN bifurcation curve SN(1,1,a), 
which will be denoted by $z^*_{\rm{sn}}$. We also note that 
the upper and lower asymmetric SN bifurcation curves form a ``horn'' 

\begin{figure}[t!]
\centerline{\epsfig{file={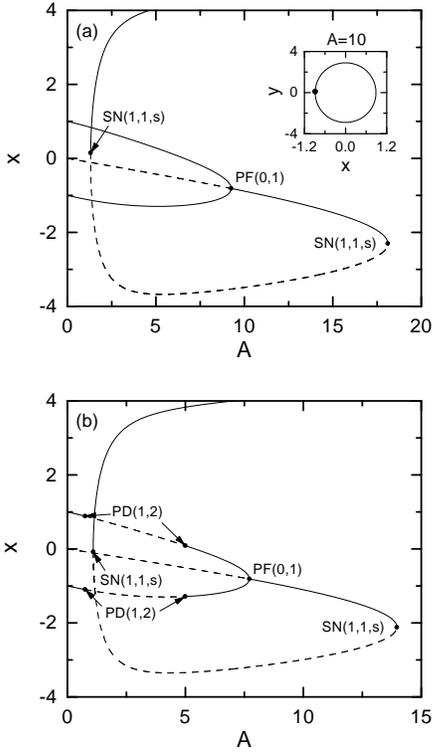}, width=5.7cm}}
\vspace{0.5cm}
\caption{Bifurcation diagrams (plots of $x$ vs. $A$) for (a) 
$\omega=3.3$ and (b) $\omega=3.0$. Here the solid line denotes a 
stable fixed point, while the dashed line represent an unstable
fixed point. The symbols denote the same as those in Fig.~1.   
The phase flow of the stabilized symmetric orbit for $A=10$ is 
denoted by a solid curve in the inset in (a), and its Poincar\'{e} 
map is represented by a solid circle. 
     }
\label{fig:BD1}
\end{figure}

\noindent{with  a cusp at $\omega = \omega_{\rm r}$, as in the asymmetric Toda 
oscillator \cite{Kurz}.}

We now present the concrete examples of bifurcations associated with 
dynamic stabilization of the symmetric saddle fixed point. The 
bifurcation diagram and the phase-flow and Poincar\'{e}-map plots are 
also given for clear presentations of the associated bifurcations. 

We first consider the case of $\omega > \omega_{\rm r}$. For 
$\omega = 2 \omega_{\rm r} [\simeq 2 \sqrt{2}]$, a subharmonic 
resonance occurs in which the asymmetric fixed points $z^*_{\rm a}$ 
become unstable by a PD bifurcation. Note that the PD bifurcation 
curve PD(1,2), belonging to the subharmonic resonance, becomes folded 
back at $\omega = \omega_{\rm f} (\simeq 3.23)$ [see Fig.~\ref{fig:SD}
(a)]. Hence, when $\omega > \omega_{\rm f}$ no PD bifurcations occur 
for the asymmetric fixed points $z^*_{\rm a}$. As an example, we 
consider the case of $\omega=3.3$. As shown in the bifurcation diagram 
in Fig.~\ref{fig:BD1}(a), the symmetric saddle fixed point $z^*_{\rm 
s}$ becomes stable through a PF bifurcation for $A=9.255\,\cdots$ by 
absorbing a pair of asymmetric fixed points $z^*_{\rm a}$. Then a 
stabilized symmetric orbit, encirling the unstable equilibrium point 
of the potential, appears, which is shown for $A=10$ in the inset in 
Fig.~\ref{fig:BD1}(a). After this dynamic stabilization, the dynamical 
behavior becomes similar to that of the single-well Duffing oscillator, 
because the ef-

\begin{figure}
\centerline{\epsfig{file={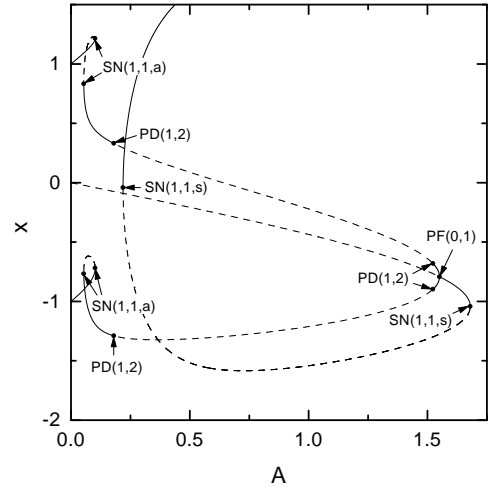}, width=6.3cm}}
\vspace{0.5cm}
\caption{Bifurcation diagram (plot of $x$ vs. $A$) for $\omega=1.2$. 
Here the solid line denotes a stable fixed point, while the dashed line 
represent an unstable fixed point. The symbols also denote the same as 
those in Fig.~1.  
     }
\label{fig:BD2}
\end{figure}

\noindent{fect of the central potential barrier on the dynamics of 
the  system becomes negligible. Such a stabilized symmetric fixed point 
also disappears for $A=18.105\,\cdots$ through a SN bifurcation by 
absorbing a symmetric unstable fixed point, born for 
$A=1.309\, \cdots$ via a symmetric SN bifurcation. }

In the range of $\omega_{\rm r} < \omega < \omega_{\rm f}$, the 
mechanism of dynamic stabilization of the symmetric saddle fixed point 
is the same, except for the bifurcation behavior of the asymmetric 
fixed points $z^*_{\rm a}$. As an example, we consider the case of 
$\omega=3.0$. As shown in Fig.~\ref{fig:BD1}(b), the asymmetric fixed 
points $z^*_{\rm a}$ lose their stability for $A=0.748\, \cdots$ by 
forward PD bifurcations, but they become restabilized for 
$A=4.991\, \cdots$ by backward PD bifurcations. Note that the 
subsequent bifurcations, associated with dynamic stabilization, are 
the same as those for the above case of $\omega=3.3$.

We next consider the case of $\omega < \omega_{\rm r}$. An example
for $\omega=1.2$ is shown in Fig.~\ref{fig:BD2}. Unlike the case of  
$\omega > \omega_{\rm r}$, the asymmetric fixed points $z^*_{\rm a}$ 
disappear for $A=0.101\, \cdots$ by the asymmetric SN bifurcations, 
and then jump phenomena occur in which the small asymmetric fixed 
points $z^*_{\rm a}$ are replaced by the relatively large asymmetric 
fixed points $z^*_{\rm {sn}}$, born for $A \simeq 0.053\,75$ via the 
asymmetric SN bifurcations. After that, the replaced asymmetric
fixed points $z^*_{\rm {sn}}$ play the same role for the dynamic
stabilization as the small asymmetric fixed points $z^*_{\rm a}$ 
do in the above case of $\omega > \omega_{\rm r}$. Hence the 
symmetric saddle fixed point $z^*_{\rm s}$ becomes stable via a PF 
bifurcation for $A=1.550\, \cdots$ by absorbing $z^*_{\rm {sn}}$.

Finally, we discuss the bifurcation behavior of the 
double-well Duffing oscillator after the dynamic stabilization of the 
symmetric saddle fixed point. We note that the stabilized symmetric
orbit, encircling the unstable equilibrium point of the potential, 
corresponds to 

\begin{figure}
\centerline{\epsfig{file={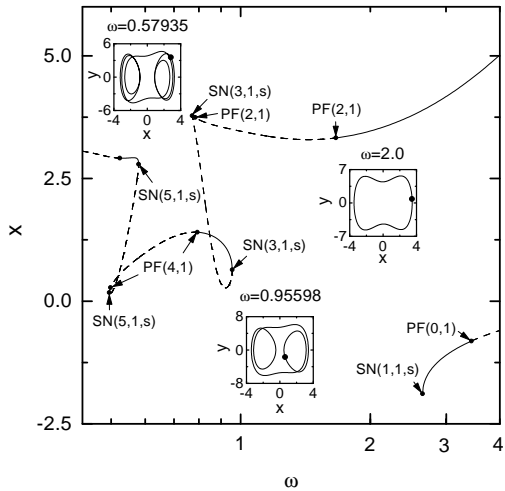}, width=6.6cm}}
\vspace{0.5cm}
\caption{Bifurcation diagram (plot of $x$ vs. $\omega$) for $A=10$. 
Here the solid line denotes a stable fixed point, while the dashed line 
represent an unstable fixed point. The symbols also denote the same as 
those in Fig.~1. The phase portraits of the three large symmeric orbits, 
born by the symmetric SN bifurcations of type (1,1,s), (3,1,s) and 
(5,1,s), are shown for $\omega=2.0$, $\omega =0.95598$, and 
$\omega=0.57935$, respectively. 
     }
\label{fig:BD3}
\end{figure}

\noindent{the symmetric stable orbit, arising from the stable
equilibrium point in the single-well Duffing oscillator. Consequently,
the double-well Duffing oscillator behaves as the single-well Duffing
oscillator, because the central potential barrier has no significant
effect on the motion of the system. As an example, we present a
bifurcation diagram (plot of $x$ vs. $\omega$) in Fig.~\ref{fig:BD3}, 
obtained by the frequency scanning for $A=10$. The symmetric orbit 
stabilized through the PF bifurcation of type (0,1) disappears for 
$\omega=2.646\, \cdots$ by a symmetric SN bifurcation of type (1,1,s), 
and then jumps to a large symmetric orbit born for $\omega=9.306\, 
\cdots$ via a symmetric SN bifurcation of type (1,1,s). A phase 
portrait of the large symmetric orbit for $\omega = 2.0$ is shown in 
the inset. As $\omega$ is decreased, large symmetric orbits with 
higher odd torsion numbers $p$ $(p=3,5,\dots)$, encircling the three 
equilibrium points of the potential, appear successively. The phase
portraits of the large symmetric orbits with torsion numbers $p=3,5$
are given for $\omega=0.955\,98$ and $0.579\,35$ in the insets, 
respectively. Note that the large symmetric orbits with higher torsion
numbers have an increasing number of loops. Furthermore, each
symmetric orbit with odd torsion number $p$ loses its stability 
through a symmetry-breaking PF of type (p+1,1). Two such PF 
bifurcations of type $(2,1)$ and $(4,1)$ are also shown in Fig.~4.
Then large asymmetric orbits with broken symmetry undergo 
period-doubling cascades. We note that all these bifurcation behaviors
are essentially the same as those in the single-well Duffing 
oscillator (refer to Fig.~1 in Ref.~\cite{Parlitz2}).}

To confirm the above numerical results, we also constructed the 
electronic analog simulator of Eq.~(\ref{eq:DDO1}) with the usual 
operational amplifiers and multipliers, and made an analog study.
Dynamic stabilization of the saddle fixed point was thus observed
experimentally. The details on the experimental results 
will be given elsewhere \cite{ytkim}.

\acknowledgments
We would like to thank W. Lim for his assistance in the numerical 
computations. This work was supported by the Korea Research 
Foundation under Project No. 1998-015-D00065 (S.Y.K.) and 
1998-015-D00060 (Y.K.) and by the Biomedlab Inc. (S.Y.K.).

\end{multicols}

\begin{references}
\bibitem{DWD} J. Guckenheimer and P. Holmes, {\it Nonlinear 
Oscillations, Dynamical Systems, and Bifurcations of Vector Fields} 
(Springer, New York, 1983), Sec.\ 2.2; F. C. Moon, {\it Chaotic and 
Fractal Dynamics} (Wiley, New York, 1992), Secs.\ 6.2 and 7.7.
\bibitem{Holmes} P. J. Holmes, Philo.\ Trans.\ R. Soc.\ London Ser.\ A 
{\bf 292}, 419 (1979).
\bibitem{Moon1} F. C. Moon and P. J. Holmes, J. Sound Vib.\ {\bf 65}, 
275 (1979); F. C. Moon, ASME J. Appl. Mech. {\bf 47}, 638 (1980).
\bibitem{Mahaffey} R. A. Mahaffey, Phy. Fluid {\bf 19}, 1387 (1976).
\bibitem{Moon2} F. C. Moon and G.-X. Li, Phys. Rev. Lett. {\bf 55}, 
1439 (1985).
\bibitem{Hop} F. T. Arecchi and F. Lisi, Phys. Rev. Lett. {\bf 49}, 
94 (1982); H. Ishii, H. Fujisaka, and M. Inoue, Phys. Lett. A 
{\bf 116}, 257 (1986).
\bibitem{IP} P. L. Kapitza, in {\it Collected Papers of P. L. 
Kapitza}, edited by D. Ter Haar (Pergamon, London, 1965), p.\ 714; 
S.-Y. Kim and B. Hu, Phys.\ Rev.\ E {\bf 58}, 3028 (1998).
\bibitem{Parlitz1} U. Parlitz and W. Lauterborn, Phys.\ Lett.\ A 
{\bf 107}, 351 (1985).
\bibitem{Gucken1} J. Guckenheimer and P. Holmes, {\it Nonlinear 
Oscillations, Dynamical Systems, and Bifurcations of Vector Fields} 
(Springer-Verlag, New York, 1983),  p.\ 24.
\bibitem{Gucken2} J. Guckenheimer and P. Holmes, {\it Nonlinear 
Oscillations, Dynamical Systems, and Bifurcations of Vector Fields} 
(Springer-Verlag, New York, 1983),  Sec.~3.5.
\bibitem{Parlitz2} U. Parlitz, Int. J. Bif. Chaos {\bf 3}, 703 (1993) 
and references therein.
\bibitem{Kurz} T. Kurz and W. Lauterborn, Phys.\ Rev.\ A {\bf 37}, 1029 
(1988).
\bibitem{ytkim} Y. Kim, S.-Y. Lee, and S.-Y. Kim (to be published).

\end{references}
\end{document}